\documentstyle[12pt]{article}
\newcommand{\rf}[1]{(\ref{#1})}
\newcommand{\beq}{\begin{equation}}
\newcommand{\eeq}{\end{equation}}

\renewcommand{\a}{\alpha}

\newcommand{\om}{\omega}

\newcommand{\bea}{\begin{eqnarray}}
\newcommand{\eea}{\end{eqnarray}}

\newcommand{\cS}{{\cal S}}

\newcommand{\cU}{{\cal U}}
\newcommand{\cG}{{\cal G}}
\newcommand{\cZ}{{\cal Z}}
\newcommand{\noi}{\noindent}

\newcommand{\oh}{\frac{1}{2}}

\newcommand{\SL}{SL(2,{\bf Z})}

\newcommand{\FSS}{Fuchs, Schellekens and Schweigert }
\newcommand{\bl}{genus one holomorphic one-point block}
\newcommand{\ver}{\thinspace\vert\thinspace}

\begin{document}
\topmargin 0pt
\oddsidemargin 5mm
\headheight 0pt
\topskip 0mm
\addtolength{\baselineskip}{0.20\baselineskip}
\pagestyle{empty}
\hfill 
\begin{center}
\vspace{3 truecm}
{\Large \bf The Untwisted Stabilizer in Simple Current Extensions}
\vspace{5 truecm}

{\large Peter Bantay}
\vspace{1 truecm}

{\em Mathematics Department\\
University of California at Santa Cruz \\}

\vspace{3 truecm} \end{center} \noi 
\underbar{\bf Abstract} A method is presented to compute  
the order of the untwisted stabilizer of a simple current orbit,
as well as some results about the properties of the resolved fields
in a simple current extension. 

\vfill \newpage
\pagestyle{plain}

In a recent paper \cite{fss}, \FSS presented an Ansatz to describe the
modular properties of a CFT obtained by simple current extensions ( for 
a review see \cite{sc} ). An important role is played in their Ansatz
by the so-called {\it untwisted stabilizer}, a subgroup of the ordinary
stabilizer defined by some cohomological properties. In general it is
quite difficult to determine the untwisted stabilizer, as it is related
to the tranformation properties of the genus one holomorphic one-point
blocks of the simple currents under the mapping class group action, and
the latter data are not readily available. 

The purpose of this note is to present a procedure that allows under
suitable circumstances the determination of the order of the untwisted
stabilizer solely from the knowledge of the $SL(2,{\bf Z})$ representation
on the space of genus one characters of the original theory. The idea is
to exploit the Frobenius-Schur indicator introduced for CFTs in \cite{FS}.

We shall not go into the details of the FSS Ansatz \cite{fss}, let's
just recall the basic setting. We are given some CFT and a group $\cG$
of integral spin simple currents, and we would like to construct the
$\SL$ representation of the new CFT obtained by extending the original
one with the simple currents in $\cG$. The first thing to
do is of course to determine the primaries of the extended theory. The
scheme is as follows :  

\begin{enumerate} 
\item First we keep only
those primaries $p$ of the original theory which have zero monodromy
charge with respect to all the simple currents in $\cG$, resulting in
a set $I_0^\cG$.  
\item In the next step we identify those primaries
from  $I_0^\cG$ that lie on the same $\cG$-orbit, i.e. for which there
exists a simple current in $\cG$ transforming one into the other.
\item We split the so obtained orbits into several new primaries.
Naively one would think that each orbit $[p]$ should be split into
$\ver\cS_p\ver$ new ones, where $\cS_p=\{\a\in \cG\ver \a p=p\}$ is
the stabilizer of the orbit $[p]$, but it turns out that this
so-called fixed point resolution 
\footnote{As this last step is the crucial part of the construction,
sometimes one refers to the extended theory  as the {\it resolved} one.}
is governed by a subgroup $\cU_p$
of the full stabilizer $\cS_p$, the so-called {\it untwisted
stabilizer} \cite{fss}. The actual definition of $\cU_p$ involves the
consideration of the space of \bl s for the simple currents in $\cG$,
thus it is in general a difficult problem to determine $\cU_p$.
\end{enumerate}

After having performed the above three steps, we are led to the
following description of the primary fields of the extended theory :
they are in one-to-one correspondence with pairs $(p,\psi)$, where $p$
is a $\cG$-orbit of $I_0^\cG$, and $\psi$ is a linear character of
$\cU_p$, i.e. an element of the dual group $\hat\cU_p$. As it cannot
lead to confusion, we shall also denote by $p$ any representative of
the orbit $p$.

Now that we know how to describe the primaries, we can formulate the
FSS Ansatz describing the $\SL$ representation of the resolved theory.
For the  exponentiated conformal weights
$\om_p=\exp(2\pi\imath\Delta_p)$ of the primaries -
 i.e. the eigenvalues of the $T$-matrix - we have \beq
 \om_{(p,\psi)}=\om_p ,\label{TM}\eeq while for the $S$-matrix the FSS
 Ansatz reads \cite{fss}

\beq S_{(p,\psi),(q,\chi)}=e_p e_q\frac{\ver\cG\ver }
{\ver\cS_p\thinspace\vert\vert \cS_q\ver}
\sum_{\a \in \cU_p\cap\cU_q}\psi(\a)S_{pq}(\a)
\bar\chi(\a), \label{SM}\eeq
where 

\beq e_p=[\cS_p:\cU_p]^{\frac{1}{2}}\label{edef}\eeq
is the square root of the index of $\cU_p$ in $\cS_p$, which is known to
be an integer on general grounds, and $S_{pq}(\a)$ is the matrix element
of the mapping class $S$ acting on the space of \bl s of the simple
current $\a$, in the canonical basis of \cite{sce}. 
In particular, $S_{pq}(0):=S_{pq}$ is just the ordinary $S$-matrix of
the original theory we started with. It may be shown 
\cite{fss,sce} that the above Ansatz leads to a consitent $\SL$
representation for the extended theory, e.g. it satisfies the relations of
\cite{mcg}.

Our aim is to determine the order of $\cU_p$, or what is the
same, the $e_p$-s from Eq. \rf{edef}. The first thing
to note is that, while to determine the actual matrix elements of $S$
through Eq. \rf{SM} we need to know all of the matrix elements
$S_{pq}(\a)$ - which are in general hard to compute -, 
upon summing over $\chi\in\cU_q$ in Eq. \rf{SM}, a standard argument 
of character theory yields

\beq \sum_{\chi\in \hat \cU_q}S_{(p,\psi),(q,\chi)}=
\frac{e_p}{e_q}[\cG:\cS_p]S_{pq}.\label{S}\eeq
This is the basic relation that we shall exploit in the sequel.

As a first corrolary of Eq. \rf{S}, by substituting for
$q$ the vacuum $0$ in the above formula,  we get that
\beq S_{0(p,\psi)}=e_p[\cG:\cS_p]S_{0p},\label{S0}\eeq
because no simple current fixes the vacuum, consequently
$\ver\cS_0\ver=e_0=1$. From this we get the following expression for the
quantum dimensions of the fields
\beq d_{(p,\psi)}=\frac{e_p d_p}{\ver \cS_p\ver}.\label{dim}\eeq

Substituting Eq. \rf{SM} into  Verlinde's formula \cite{ver}
\beq N_{pq}^r=\sum_s\frac{S_{ps}S_{qs}\bar S_{rs}}{S_{0s}}\label{verl}\eeq
and exploiting Eq. \rf{S} leads to

\beq \sum_{\psi \in \hat \cU_p,\chi\in\hat \cU_q }
N_{(p,\psi),(q,\chi)}^{(r,\rho)}={e_r\over e_p e_q}\thinspace
\frac{1}{\ver \cS_r\ver}\sum_{\alpha\in \cG}N_{pq}^{\alpha r}.\label{fuz}\eeq

We are practically done, all that is left is to
sustitute Eqs. \rf{TM}, \rf{S0} and \rf{fuz} into the 
formula for the Frobenius-Schur indicator \cite{FS}

\beq \nu_p=\sum_{q,r}N_{qr}^pS_{0q}S_{0r}\frac{\om_q^2}{\om_r^2}
\label{FSdef}, \eeq
to arrive at

\beq \nu_{(p,\psi)}=\frac{e_p}{\ver \cS_p\ver}\sum_{\alpha\in \cG}
{\cal Z}[p,\alpha],\label{FS}\eeq
where ${\cal Z}[p,\alpha]$ stands for the $\cZ$-matrix introduced
in \cite{FS}, with matrix elements

\beq {\cal Z}[p,q]=\om_q^{-\oh}\sum_{r,s}N_{rs}^pS_{qr}S_{0s}
\frac{\om_s^2}{\om_r^2}.\label{Zdef}\eeq

If we  introduce the notation 
\beq {\cal Z}_\cG(p):=\frac{1}{\ver \cS_p\ver}\sum_{\alpha\in \cG}
{\cal Z}[p,\alpha],\label{ZG}\eeq
then because the lhs. of Eq. \rf{FS} can only take the values $\pm 1$ and
$0$, we get the following alternative : 

\begin{enumerate}
\item ${\cal Z}_\cG(p)=0$. In this case the
resolved field $(p,\psi)$ is complex, and we get no information on
$e_p$ from Eq. \rf{FS}.

\item ${\cal Z}_\cG(p)\ne 0$. In this case
the resolved field $(p,\psi)$ is either real or pseudo-real according
to the sign of ${\cal Z}_\cG(p)$, moreover
\beq e_p=\frac{1}{\ver{\cal Z}_\cG(p)\ver},\label{ramindex}\eeq
or in other words 
\beq \ver\cU_p\ver=\ver\cS_p\ver{\cal Z}_\cG^2(p).\label{Uorder}\eeq
\end{enumerate}
We see that at least for some of the primaries $p$ we can determine
$e_p$ from the knowledge of  ${\cal Z}_\cG(p)$, which is in turn
completely determined by the $\SL$ action on the space of genus one
characters of the original theory.

What to do for those primaries for which ${\cal Z}_\cG(p)=0$ ? Well, in
that case there is still some hope to determine $e_p$ through a
similar procedure, because Eq. \rf{FS} is just a special case of the
more general relation

\beq \sum_{\chi\in \hat \cU_q}{\cal Z}\left[(p,\psi),(q,\chi)\right]=
\frac{e_p}{e_q}\frac{1}{\ver \cS_p\ver}\sum_{\alpha\in \cG}
{\cal Z}[p,\alpha q].\label{Z}\eeq
Exploiting the properties of the $\cZ$-matrix \cite{FS}, namely that its matrix
elements ${\cal Z}[p,q]$ are integers of the same parity as $N_{pp}^q$ 
and bounded in absolute value by the latter, sometimes it is possible
to get from Eq. \rf{Z} the index $e_p$ for primaries for
which Eq. \rf{FS} does not yield an answer.

In summary, we have seen that a great deal of information about the
extended theory may be learned simply by looking at the $\SL$ action on
the space of characters of the original theory, e.g. in some cases the
indices $e_p$ may be determined without having to look at the mapping
class group action on the space of \bl s. While the knowledge of
the $e_p$-s is not enough in general to determine $\cU_p$ itself,
it is nevertheless enough for the correct counting of the
primary fields of the resolved theory and the computation of their
quantum dimensions and other characteristics. Moreover, in many
practical instances $e_p$ completely determines the untwisted
stabilizer, e.g. in the obvious cases $e_p=1$ and $e_p^2=\ver\cS_p\ver$.

\vspace{1.5 truecm}

{\it Acknowledgement :} It is a pleasure to acknowledge discussions with 
Geoff Mason and Christoph  Schweigert.

\vspace{2 truecm}

Supported partially by OTKA T016251

e-mail: bantay@hal9000.elte.hu

\end{document}